# Coronal Imaging with the *Solar UltraViolet Imager*

Sivakumara K. Tadikonda[1], Douglas C. Freesland[2], Robin R. Minor[3], Daniel B. Seaton[4, 5], Gustave J. Comeyne[6], Alexander Krimchansky[7]

[1]CSEngineering, Annapolis, MD, USA, [2]ACS Engineering, Clarksville, MD, USA
[3]ASRC Federal Technical Services, MD, USA
[4]Cooperative Institute for Research in Environmental Sciences, Univ. Of Colorado, Boulder, CO, USA
[5]NOAA National Centers for Environmental Information, Boulder, CO, USA
[6]National Oceanic and Atmospheric Administration, Silver Spring, MD, USA
[7]NASA Goddard Space Flight Center, Greenbelt, MD

*sivakumara.k.tadikonda@nasa.gov* , *douglas.c.freesland@nasa.gov* , *robin.r.minor@nasa.gov* , *daniel.seaton@noaa.gov* , *gustave.j.comeyne@nasa.gov* , *alexander.krimchansky@nasa.gov*

**Abstract**

We investigate the coronal imaging capabilities of the *Solar UltraViolet Imager* (SUVI) on the *Geostationary Operational Environmental Satellite-R* series spacecraft. Nominally Sun-pointed, SUVI provides solar images in six Extreme UltraViolet (EUV) wavelengths. On-orbit data indicated that SUVI had sufficient dynamic range and sensitivity to image the corona to the largest heights above the Sun to date while simultaneously imaging the Sun. We undertook a campaign to investigate the existence of the EUV signal well beyond the nominal Sun-centered imaging area of the solar EUV imagers. We off-pointed SUVI line-of-sight by almost one imaging area around the Sun. We present the details of the campaign conducted when the solar cycle is at near the minimum and some results that affirm the EUV presence to beyond three solar radii.

**1 Introduction**

In general, the solar corona is routinely observed above about 1.7 solar radii using white-light coronagraphs or during eclipses, but has rarely been studied in extreme ultraviolet (EUV) at these large heights. It is widely believed that almost all EUV flux from the corona is the result of collisional excitation (Golub and Pasachoff, 2010) and is proportional to the square of the density of the emitting plasma. As a result, EUV emission is expected to fall off rapidly enough that the signal from large heights above the heliosphere where coronal density is low is too weak to be detected by EUV imagers. Observations of the corona at large heights have been routinely made for decades using the *Large Angle Spectroscopic Coronagraph* (LASCO; Brueckner et al. 1995) on the *Solar and Heliospheric Observatory* (SOHO) spacecraft. However, LASCO's inner coronagraph, C1, failed after only a short period of time, restricting the field of view of the instrument to heights above about 2.2 solar radii, leaving an observational gap between the solar surface and the extended corona that, to date, has not been routinely filled by any alternative observations.

A few instrument studies have reported observations of the EUV corona at larger heights; for example, the SPIRIT telescope on board CORONAS-F (see, for example, Slemzin et al. 2008), the *Telescope-Spectrometer for Imaging Solar Spectroscopy* (TESIS; Kuzin, 2011) imager on the Russian *Complex Orbital Observations Near-Earth of Activity of the Sun-PHOTON* (CORONAS-PHOTON)*,* and the *Sun Watcher with Active Pixels and Image Processing* (SWAP; Seaton et al. 2013a, Halain et al. 2013) on the European Space Agency's *Project for On-Board Autonomy 2 (*PROBA2*)* spacecraft. Seaton et al. (2013b) reported on the evolution of the corona at 17.4 nm on long timescales at heights above 1.6 solar radii using SWAP and demonstrated that large, fan-like structures that extend well beyond the SWAP field of view appear to dominate EUV emission in this region. Goryaev et al. (2014) reported in detail on observations of one such structure using off-pointed SWAP observations, in which they could track at least some emission from the structure to heights as large as almost 2.5 solar radii. Reva et al. (2017) tracked an eruption in the 17.1 nm channel of TESIS to heights as great as 2 solar radii, while D'Huys et al. (2017) used multi-perspective observations from SWAP and the *Sun Earth Connection Coronal and Heliospheric Investigation* (SECCHI; Howard 2008) instrument on NASA's *Solar-Terrestrial Relations Observatory (STEREO)* spacecraft to continuously track the acceleration of an eruption through heights greater than two solar radii.

At the same time, observing the low corona — below two solar radii — in white light is also difficult, due to stray light that arises from the extremely bright photosphere, even when well-occulted. This is particularly true for space-based coronagraphs, where the distance between occulter and imager is strongly limited by the rocket fairing constraint considerations. One approach to this problem is formation-flying instruments with separation baselines >100 m, like the planned PROBA3 spacecraft (Shestov and Zhukov, 2018). However, since such formational-flying is possible only for a limited duration and over a small portion of an orbit, these instruments cannot be used operationally to track coronal mass ejection (CME) initiation.

Thus, a growing body of evidence suggests that exploring the EUV corona at much larger heights than have previously been explored is not only worthwhile, but also likely to yield important observations, particularly the mechanisms that accelerate and shape coronal eruptions early in their evolution. Observations from the *Solar UltraViolet Imager* (SUVI) on National Oceanic and Atmospheric Administration (NOAA)'s *Geostationary Operational Environmental Satellite-16 (GOES-16)* of the powerful solar eruption on 2017 September 10 (Seaton and Darnel, 2018) showed the instrument could easily track an eruption to heights as great as 1.7 solar radii, and likely to much greater heights if the field of view allowed it. Such observations are of value not only because they provide valuable data that can help improve our understanding of CME initiation, but also because they provide valuable operational data. Early CME acceleration profiles can be of use in forecasting the kinematics of a CME and play a key role in determining the intensity of eruptional-associated solar energetic particle events (see, for example, Gopalswamy et al. 2018).

We set out to study the viability of using EUV images to: 1) characterize the corona at heights extending from the solar surface out to a few solar radii, 2) track CMEs from their origins over these heights and, 3) after collecting and analyzing images in 17.1 nm and 19.5 nm wavelengths, investigate whether one is clearly preferable for capturing CMEs. We off-point SUVI from the Sun to collect image data. In this paper, we describe the details of a test campaign, present results, and discuss the degree of our success in achieving the stated objectives. We also discuss how such observations using SUVI and other proposed future instruments could improve our understanding of coronal physics, and how such a campaign could serve as one of several backup tools for space weather forecasting in the event that LASCO observations become unavailable.

## 2  Instrumentation and Observation

### 2.1 THE INSTRUMENT

SUVI, shown in Figure 1, is a normal-incidence Cassegrain telescope that shares considerable design heritage with the *Atmospheric Imaging Assembly* (AIA; Lemen et al. 2012) on NASA's *Solar Dynamics Observatory* (SDO) spacecraft. SUVI images the Sun in six EUV wavelengths: 9.4, 13.1, 17.1, 19.5, 28.4 and 30.4 nm (Martinez-Galarce, et al., 2010, 2013). The instrument consists of the main imaging telescope, a Guide Telescope (GT) and a Camera Electronics Box (CEB) mechanically integrated to the telescope, and a SUVI Electronics Box (SEB). The SEB provides power and data interfaces to the spacecraft. The optical chain in the telescope consists of thin film entrance filters vapor deposited on a metallic mesh, multi-layer coated primary and secondary mirrors (Martinez-Galarce, et al., 2010, 2013), a set of thin film analysis filters in two filter wheels near the focal plane, and a charge-coupled device (CCD) detector. The CCD consists of 1280x1280 pixels with a plate scale of 2.5 arcsec, and, together with the optical system, provides a nominal 53 arcminute square field-of-view (FOV) from the geostationary orbit. An aperture selector with a 60° opening, and two multi-segmented mirrors, enables the imaging in any of the six wavelengths in a single telescope body. A nominal 4-minute cadence provides for the observation of the Sun in all wavelengths while meeting large dynamic range requirements on single- and multi-spectral images. The GT provides Sun-pointing knowledge. In the nominal Sun-pointing case, the spacecraft control system uses the GT data to control the line-of-sight (LOS) to the Sun.

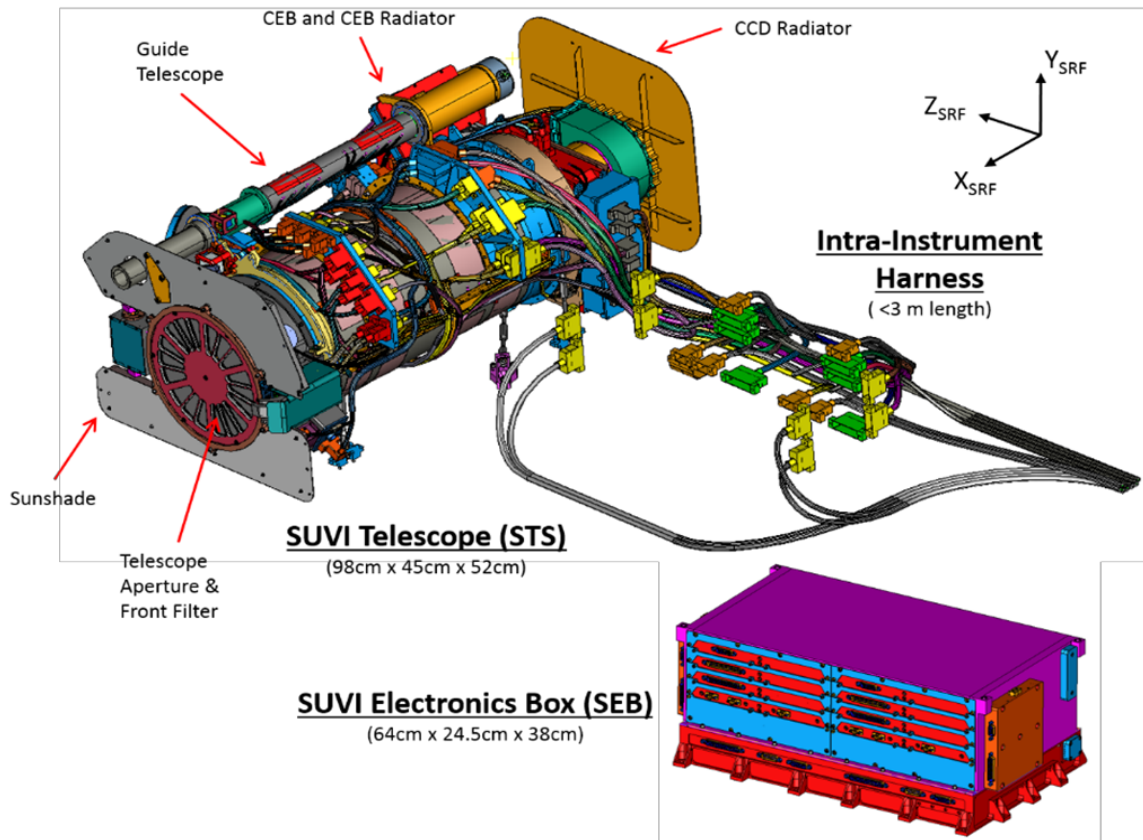

*Figure 1: Solar Ultra Violet Imager (SUVI)*

## 2.2 THE SATELLITE

GOES-16 and GOES-17, part of the GOES-R series, were launched on November 19, 2016 and March 1, 2018, respectively. GOES-16 is stationed at 75.2º W longitude and GOES-17 is at 137.2º W longitude. Figure 2 shows the satellite architecture. Each spacecraft accommodates terrestrial and space weather instruments. The nadir-pointed instruments are mounted on an Earth-Pointing Platform (EPP) and the solar instruments are located on the Sun-Pointing Platform (SPP). The EPP hosts the *Advanced Baseline Imager* (ABI) and the *Geostationary Lightning Mapper* (GLM) while the solar instruments SUVI and *EUVS XRS Irradiance Sensors* (EXIS) are mounted on the SPP. Not shown in Figure 2 explicitly is the *Space Environment In-Situ Suite* (SEISS), consisting of five sensors that are distributed on the spacecraft bus, which detects energetic particles. A boom-mounted *Magnetometer* (MAG) provides information about the local magnetic environment. ABI and GLM observe terrestrial weather while SUVI, EXIS, SEISS, and MAG provide space weather information.

The satellite is nominally Earth-pointed. The Sun-pointing for the solar instruments is achieved by the two-axis gimballed SPP. The Solar Array Drive Assembly (SADA) and the SPP Elevation Gimbal Assembly (SEGA) enable azimuth and elevation control. The spacecraft uses the GT Sun-pointing data to provide accurate and stable pointing control of the line-of-sight (LOS) to the Sun for SUVI and EXIS.

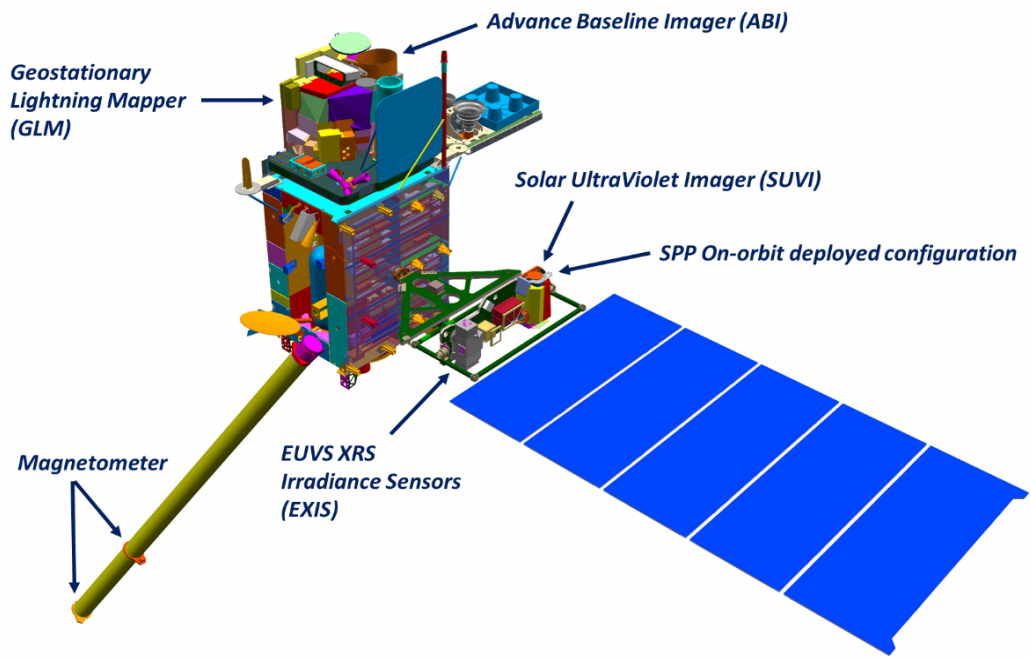

*Figure 2: GOES-R satellite architecture*

**2.3 THE OBSERVATION**

Of the six SUVI channels, the 19.5 nm channel was selected for demonstrating the instrument's imaging capability as shown in Figure 3, composed with the images from the SUVI on GOES-16. The large coronal structure is evident and it demonstrated that SUVI possesses significant signal-to-noise ratio far above the solar disk. Of the remaining channels (figures shown at the bottom in Figure 3), only 17.1 nm routinely reveals significant signal in the extended corona, although occasional observations — particularly during strong solar eruptions and flares — have revealed extended structures in other wavelengths. Together with the SUVI observations of the 2017 September 10 X8.2 flare (Seaton and Darnel, 2018), the idea of investigating the presence of the EUV corona outside the nominal SUVI FOV was born.

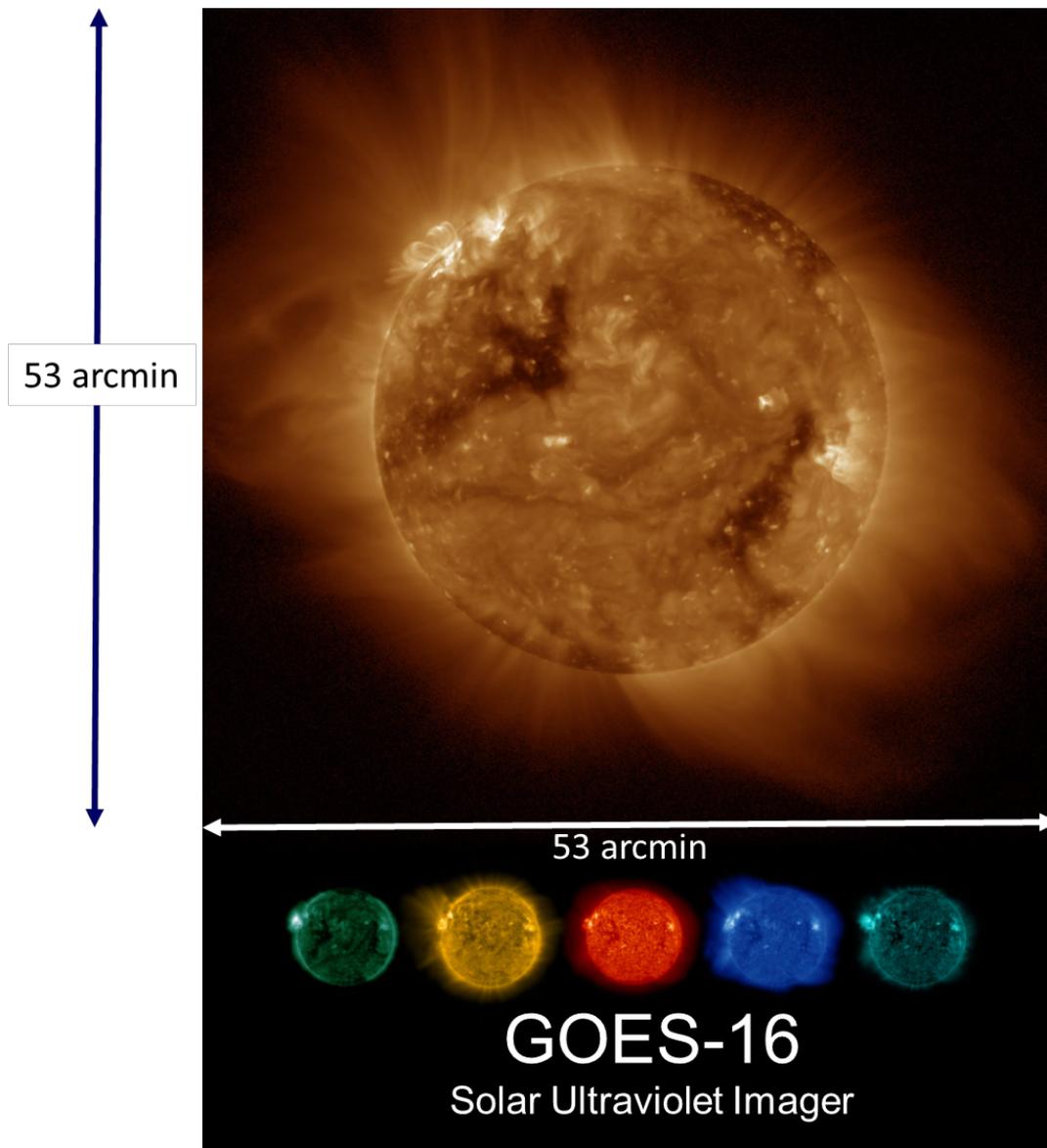

*Figure 3: Early observations of the corona in six wavelengths from SUVI on GOES-16. Large image: 19.5 nm; Bottom left to right: 9.4 nm, 17.1 nm, 30.4 nm, 28.4 nm, and 13.1 nm*

**2.4 THE APPROACH**

On-orbit SUVI calibration tests necessitating the movement of the SUVI line-of-sight off the Sun demonstrated that the resulting dynamic disturbances were compliant with the requirements set for the Earth-pointing instruments. Thus, SUVI maneuvers could be carried out without significantly disrupting Earth-observing operations. Taking advantage of the excellent performance of the spacecraft attitude and gimbal control systems, we conducted tests to assess the performance of SUVI as an extremely wide-field EUV coronal imager by off-pointing the SPP around the Sun and synthesizing a mosaic image that is about 4 times the diameter of the Sun.

We pursued three options shown in Figures 4-6.

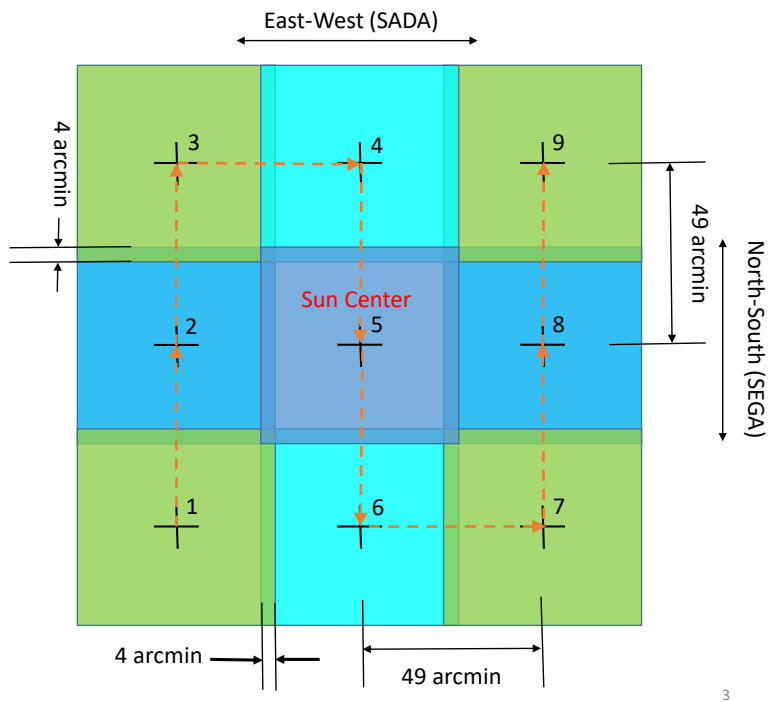

*Figure 4: Nine-panel option*

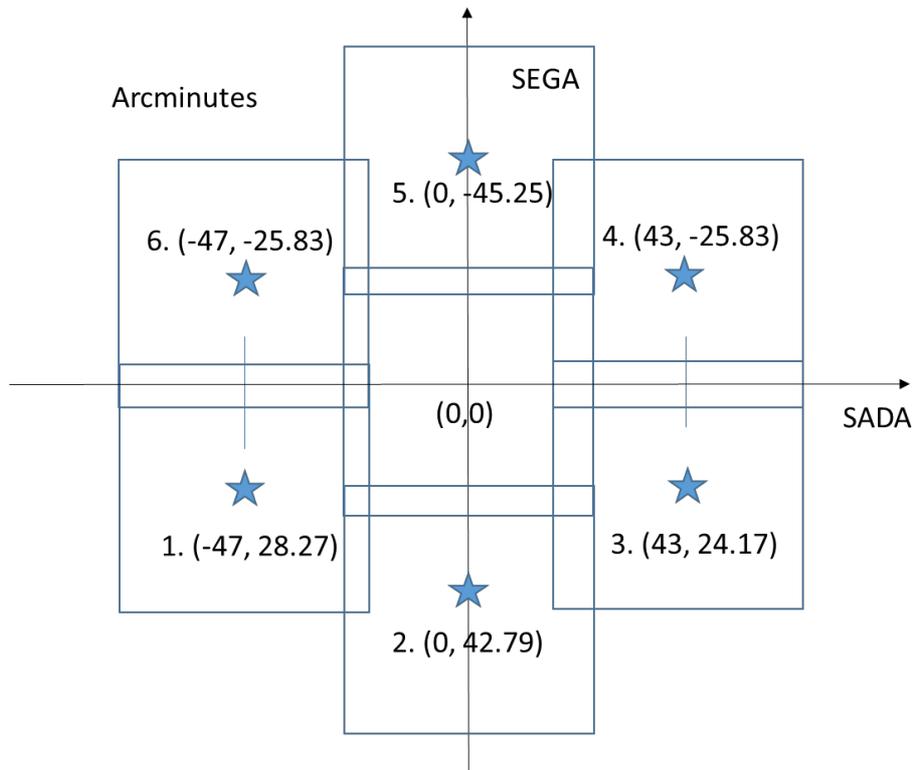

*Figure 5: Seven-panel option*

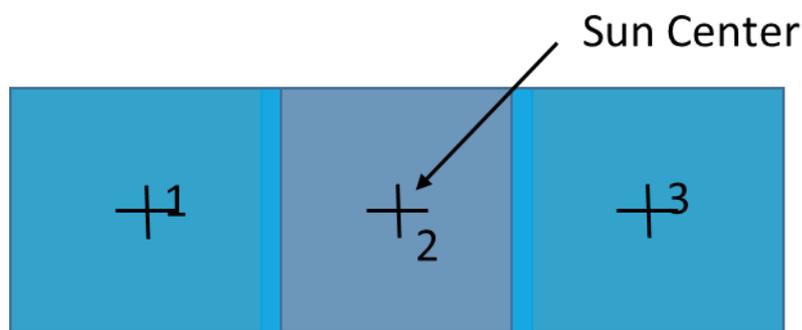
*Figure 6: Three-panel option*

In Option 1 (Figure 4), starting at the nominal Sun center indicated in the center panel, the LOS is offset to the center of panel 1. After the imaging at that location, the LOS is moved through the locations 2 to 9, in sequence. The approach for the other options is similar. The overlap seen in these figures enables navigation and accounts for potential pointing errors when composite images are synthesized.

Option 1 will provide the same area all around the Sun irrespective of seasonal variations, but — because it requires nine independent observations — it also has the slowest refresh rate. Because SUVI is aligned to celestial north rather than solar north, Option 2 (Figure 5) needs adjustment in order to address the solar P-angle change over the course of the year, which varies by ±26.3°. Seasonal solar P-angle variation will have an impact in Option 3 (Figure 6) as well, but this option is the simplest to execute and provides the fastest overall imaging cadence.

## 3 Implementation

The test campaign ran in three phases and included two spacecraft, GOES-16 and GOES-17. In the February 2018 execution on GOES-16, we tested the imaging patterns shown in Figures 4 and 5. We varied the exposure durations in the 20-200 secs range in order to better characterize the signal-to-noise ratio (S/N) so that an optimal exposure duration could be determined. For Option 1, we collected the 17.1 nm and 19.5 nm images in the forward and reverse paths, respectively, whereas, for Option 2, we collected images in both channels at each off-point. While accurate Sun-pointing is achieved using the GT data in nominal operations, for the range of off-points considered here, owing to the limited linear range of GT, a less accurate Sun sensor was necessarily employed. Consequently, larger pointing errors and slightly more jitter were expected. We allowed a 4 arcmin overlap for the adjacent panels to account for these pointing errors and enable registration for the composite images. The execution for each option lasted a couple of hours.

SUVI's focal plane filter assembly includes a glass filter, transparent to visible light but opaque to the EUV, that allows us to characterize the amount of visible light contamination in any channel. In normal operations, we anticipate a small amount of visible light contamination from the slight transparency of the zirconium foil entrance filters used for the 9.4 and 13.1 nm channels, as well as from possible pinholes in the entrance filter for any channel. However, since SUVI was not designed or characterized for off-pointed operations of the magnitudes attempted in this campaign, a potential for stray light contamination during these tests exists. We therefore collected images with the glass filters in place to assess whether visible light scattering effects due to glint could contaminate off-pointed images. However, no such scattering effects were observed and moving forward, we dropped the glass images entirely. It is worth noting that other sources of stray light, presumably internal to the telescope and in the EUV, are present depending on the channel and pointing position, and require special processing to remove their effects from the observations.

The results for Options 1 and 2 are shown in Figures 7 through 10. Figures 7 and 8 present unprocessed stitched mosaics for Option 1 and demonstrate the presence of corona to a few solar radii. However, they also show channel-specific glint – predominantly in the center-right and bottom-center panels for 17.1 nm and 19.5 nm, respectively – that is overwhelming the coronal features. Channel-specific vignetting effects can also be seen and these could be accounted for in composing better composites in future campaigns by creating larger overlaps in the appropriate directions. Figures 9 and 10 show mosaics for Option 2 that have been corrected for as much stray light as possible. To do this, we generate an image of the low frequency background by computing the minimum value in the 30×30 neighborhood for each pixel and subtract this from the coronal image. The results are subsequently scaled using an azimuthally varying radial filter (see discussions of similar filtering techniques in Seaton and Darnel, 2018; Seaton et al. 2013b) to reduce the large dynamic range falloff of the corona at large heights and improve image contrast. The techniques we used somewhat amplify the brightness of the seams between the different exposures and tend to over-correct the contrast of the disk image, but help considerably in enhancing the visibility of the extended corona. Efforts to improve this technique and remove these artifacts are presently underway.

Coronal structure is clearly visible out to a few solar radii in the 7-panel images in Figures 9 and 10. A signal-to-noise analysis of bright structures in these images showed that noise floor is dominated by the residual stray light that is not fully removed by this method, rather than detector noise or signal shot noise, thus it is not clear longer exposures would necessarily improve the image quality. Instead, a more significant gain could be achieved by improving our stray-light removal techniques. Nonetheless, a study of the noise as a function of height revealed that at 2 solar radii, the S/N ratio is roughly 10, while at 3 solar radii it is closer to 2. This is likely insufficient for high quality photometric analysis of the data, but because the eye can easily follow coherent structures in the images, even noisy ones, it is sufficient for morphological studies and use in space weather operations in which characterizing the dynamics of the corona is more important than the coronal brightness.

Likewise, analysis of the evolution of the EUV corona on long timescales (Seaton et al. 2013b) suggests the EUV brightness of the corona at large heights can vary significantly — by up to an order of magnitude — over the course of the solar cycle. Thus, the detectability of coronal signal at these heights could also change significantly depending on the point during the solar cycle at which the campaign is carried out. In this case, run under quiet conditions, the results of this test likely establish a lower limit on our ability to observe the EUV corona at large heights, rather than make a definitive statement on the farthest extent at which the EUV corona can be observed.

No CMEs occurred during this test campaign and thus we could not answer the question about selecting only one of the 17.1 and 19.5 nm channels for implementation in a routine operation. Both passbands, however, showed enough promise to warrant their inclusion in subsequent studies. These data sets also indicated that bright and extended coronal structures are generally distributed in the East-West direction and we concluded that the refresh rate for the composite image could be significantly improved by pursuing the three-panel option. It is perhaps worth noting that, in principle, the evolution of the corona from its current solar minimum configuration to solar maximum conditions, during which the EUV structure is distributed more uniformly around the solar disk (see Seaton 2013b), could necessitate a change in operations to obtain a more extended view of the corona near the poles, but a three-panel view is apparently sufficient at solar minimum.

An early S/N analysis of this phase of the testing suggested that an exposure duration of 80 s is sufficient, and the cadence can be improved by exploiting the 2x2 on-chip binning capability of the CCD for the off-pointed panels. In this case, the light gathered in a 20-s exposure is effectively equal to that collected in an un-binned 80-s exposure. We used 20-s un-binned exposures for Sun-centered images. Although this resulted in saturation of the brightest pixels inside the solar disk, we deemed this to be not significant due to our interest in capturing the details of the off-limb corona. In addition, since SUVI's CCD uses anti-blooming technology, saturation is isolated to just the few brightest structures on the disk.

We also concluded that the pointing accuracy for the off-pointed panels was better than expected and we reduced the overlap to 2 arcmin. This resulted in larger field of regard. The images also indicated no deterioration in pointing stability.

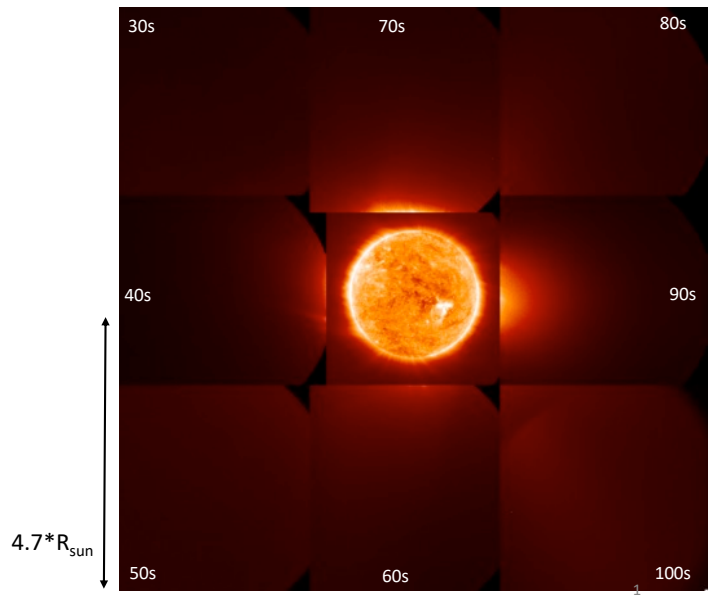

*Figure 7: 17.1 nm composite (GOES-16 Preliminary, Non-Operational Data)*

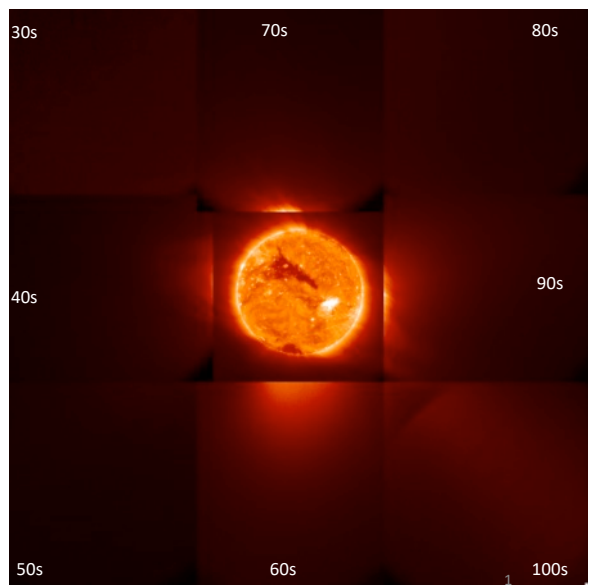

*Figure 8: 19.5 nm composite (GOES-16 Preliminary, Non-Operational Data)*

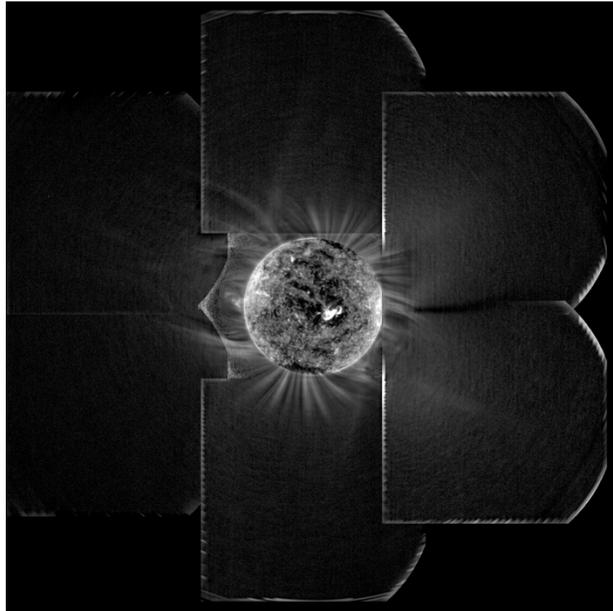

*Figure 9: Seven-panel background-subtracted composite for 17.1 nm (GOES-16 Preliminary, Non-Operational Data)*

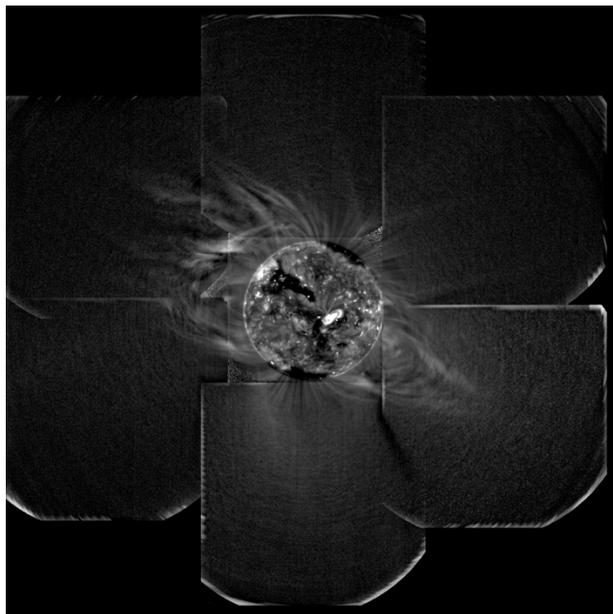

*Figure 10: Seven-panel background-subtracted composite for 19.5 nm (GOES-16 Preliminary, Non-Operational Data)*

From a systems perspective, the LOS off-pointing from the Sun is similar to the SUVI and EXIS calibrations that require gimbal movements. During these calibrations, the data indicated that SADA articulation provided the most disturbance to GLM. Although within requirements, GLM's sensitivity caused us to improve the SUVI coronal imaging scheme in order to mitigate this impact. In April 2018, we characterized the dynamic disturbances due to moving the SADA to simulate the execution of Option 3 on GOES-17 and in addition investigated the means to mitigate GLM sensitivity by varying instrument parameters. Using the data from these evaluations, we updated both the GLM imaging and the SADA slew control parameters for the follow-on phases of the testing for an optimal system performance.

With the optimal parameters selected, we conducted a 72-hour-long Extended Coronal Imaging (ECI) of the Sun starting on 2018 June 4 on GOES-17. The execution consisted of collecting images in both channels at each off-point and Sun center (Option 3), in the 1-2-3-1-2-3 … order shown in Figure 6. Figures 11 and 12 show composite images from this campaign. (These images have not been processed with the glint removal algorithm discussed earlier.) Figure 12 shows the evolution of an active region over the West limb of the Sun during the ~34-hour duration to almost the center of the right panel for the 19.5nm channel. The nomenclature for the time stamp for the images is yyyymmdd_hhmmss. Although evolution can be seen in Figure 11 for the 17.1 nm channel, it is not as extended as that for the 19.5nm channel.

A comparison of Figures 7 and 11 clearly indicates that the stray light effects for the 17.1 nm channel are seen in different locations: in the right panel in Figure 7 and left panel in Figure 11. This is due to the different spacecraft orientations in the geostationary orbit. The GOES-16 data from February 2018 is for the nominal "upright" configuration of the spacecraft in which the solar array is located below the orbital plane while the GOES-17 was in the "inverted" configuration in which the solar array is located above the orbital plane. We were fortuitous for the GOES-17 configuration for our June 2018 campaign; otherwise, the streamers in the 17.1 nm channel would have had a much lower signal-to-noise ratio and could have been dwarfed by the scattered light. It is also possible that the smaller overlap in the panels (2 arcmin vs 4 arcmin employed earlier) means a larger offset for the June 2018 campaign than that in the February 2018 case resulted in larger amount of stray light. However, considering that this data is from two different flight instruments, it is clear that two flight models of SUVI demonstrated very similar stray light characteristics for the large off-points.

For the next phase, we executed the ECI option shown in Figure 6 over a longer period in order to capture at least one full solar rotation. This campaign started on August 6, 2018 and ended on September 13, 2018. We are currently analyzing this data.

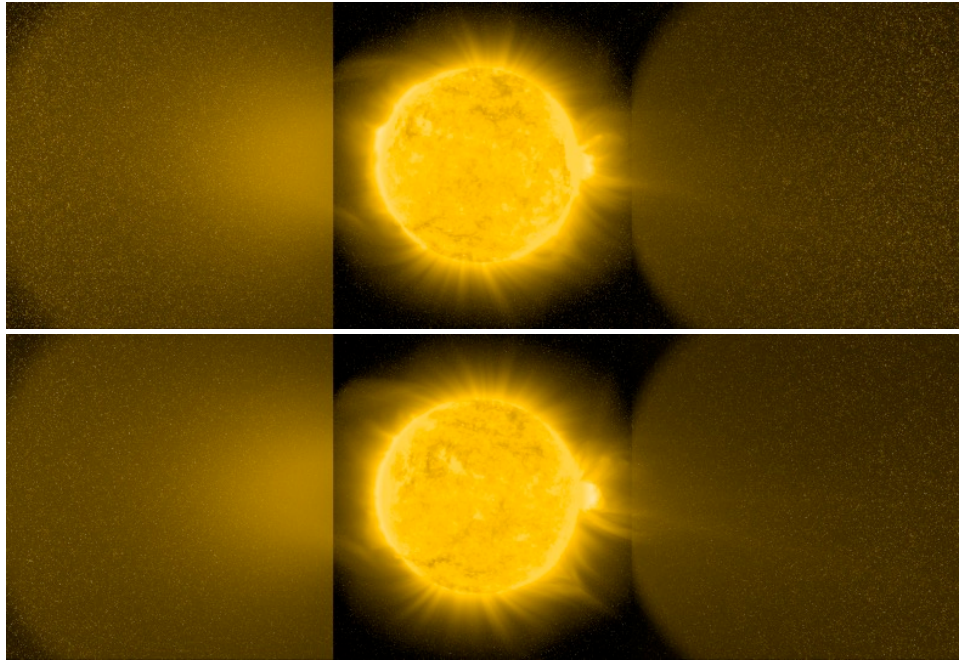

*Figure 11: ECI Phase 2 results for 17.1nm. The time stamps for top and bottom are 20180604_131458 and 20180605_230203, respectively. GOES-17 Preliminary, Non-Operational Data*

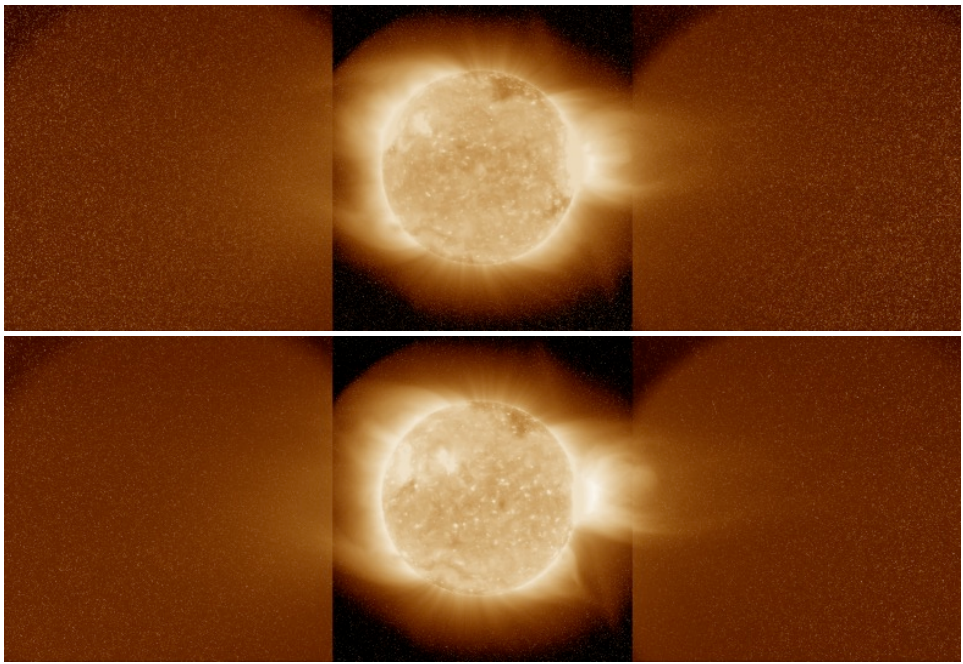

*Figure 12: ECI Phase 2 results for 19.5nm. The time stamps for top and bottom are 20180604_131523 and 20180605_230228, respectively. GOES-17 Preliminary, Non-Operational Data.*

**4 Some thoughts for future work**

The quiet part of the current Solar cycle did not provide us the CMEs that would have helped us in the selection of one of the channels if we were to choose only one. Thus, one of the objectives of this campaign has not been achieved. We believe an ECI campaign over an extended duration and with some help from solar activity can address this. GOES-16 at 75.2° W longitude is GOES-East. GOES-17 at 137.2° W longitude is GOES-West. Thus, there are two SUVI instruments pointed at the Sun. The data collected so far indicates that SUVI can function as an EUV coronagraph up to the inner edge of the occulted regions of the existing white light coronagraphs. Complete system analysis that includes the impact to EXIS products is expected to provide the information necessary to enable NOAA to determine if pursuing an approach in which either one of the SUVIs on GOES-E and GOES-W is in a continuous ECI mode is feasible.

Although this campaign clearly demonstrated the promise of extended coronal imaging with SUVI, it is not without challenges. Most importantly, as the figures demonstrated, artifacts — stray light and image spikes due to energetic particle hits — can obscure the relatively faint EUV coronal signal at large heights. Relatively effective techniques exist that can remove spikes, but stray light is more challenging. Separating stray light from the underlying coronal signal we aim to measure is not trivial, and although early analyses have shown promise, complete removal of stray light requires further work. A key step in transitioning these experimental campaigns to a routine capability will be the development of robust automated algorithms that can generate high-quality, cleaned observations that reveal the corona and enable determination of the CME speeds.

**5 Conclusion**

We demonstrated the capability to perform extended coronal imaging with the EUV instruments at the 17.1 nm and 19.5 nm wavelengths by off-pointing the SUVI boresight to locations around the Sun and generating a composite coronal image. We exploited the GOES-R series satellites' ability to provide stable, fine pointing even during the slews. We optimized the ECI pattern, exposure durations, and slew parameters during the course of the campaign using the SUVI instruments on two satellites: GOES-16 and GOES-17. The results, obtained during a relatively quiet part of the solar cycle, are promising. We did not anticipate stray light impacts but have implemented a method for enhancing the image contrast to demonstrate that coronal features are present in EUV to beyond three solar radii. Further work on this front is necessary for its automation. The GOES-16 SUVI capture of the September 10, 2018 X8.2 flare demonstrated the instrument's dynamic range and sensitivity. The ECI campaign demonstrated that if such a flare or CME had occurred during the ECI campaign, it would have been easily detected in the off-pointed images.

The routine use of an ultraviolet solar imager as both an imager and an EUV coronagraph would be the first of its kind and is an exciting development. The on-orbit solar coronagraphs to date provide white light images with inner fields of view that do not extend below 2 solar radii. SUVI observations in this region confirm that imaging in the EUV is a viable approach to studying the corona — particularly in the observational gap above heights where it is observed by traditional

EUV imagers and below the heights where space-based coronagraph observations are routinely available. The evidence from these early tests confirms a dedicated ECI campaign, with improved image processing techniques that can be developed using observations from our existing campaigns, could provide valuable complementary observations that would enhance space weather forecasting capabilities. Likewise, SUVI observations such as these provide a pathfinder dataset for a variety of proposed instruments such as the *Coronal Spectrographic Imager* (COSIE; Savage, 2016) or the *Sun's Coronal Eruption Tracker* (SunCET; Chamberlin, 2018) cubesat.

**Acknowledgements**

The authors sincerely thank the GOES-R Flight Project for the test campaign, and gratefully acknowledge the Lockheed Martin (LM), Palo Alto, CA, SUVI team's assistance in the campaign and the analysis. Special thanks to Margaret Shaw, Lawrence Shing, and Ralph Seguin of LM, and Calvin Nwachuku of the GOES-R Mission Operations Support Team for their assistance in this effort.